\documentclass[aps,pra,twocolumn,showpacs,floatfix]{revtex4}

\usepackage{bm}
\usepackage{epsfig}

\newcommand{\p}{\partial}
\newcommand{\Topo}{{\cal N}}
\newcommand{\topo}{n}
\newcommand{\lex}{\ell_{\rm ex}}
\newcommand{\half}{\frac{1}{2}}
\newcommand{\winding}{S}

\begin{document}

\title{Rotating vortex dipoles in ferromagnets}
\author{S. Komineas}
\affiliation{Max-Planck Institute for the Physics of
Complex Systems, N\"othnitzer Str. 38, 01187, Dresden, Germany.}

\date{\today}

\begin{abstract}
Vortex-antivortex pairs are localized excitations and have been found to be
spontaneously created in magnetic elements.
In the case that the vortex and the antivortex have opposite polarities the
pair has a nonzero topological charge, and it behaves as
a rotating vortex dipole.
We find theoretically, and confirm numerically,
the form of the energy as a function of the angular momentum of the system
and the associated rotation frequencies.
We discuss the process of annihilation of the pair
which changes the topological charge of the system by unity
while its energy is monotonically decreasing.
Such a change in the topological charge affects profoundly the
dynamics in the magnetic system.
We finally discuss the connection of our results
with Bloch Points (BP) and the implications for BP dynamics.
\end{abstract}

.
\maketitle

\section{Introduction}
\label{sec:introduction}

Magnetic vortices have a nontrivial topology which is responsible
for their stability and
for the fact that they are central in theoretical studies for the
micromagnetic description of
two-dimensional (2D) magnets and thin films \cite{hubert}.
Similar topological structures arise in many physical systems.
In particular, magnetic vortices have been discussed extensively
in quantum field theory \cite{rajaraman}.
However, experimental evidence for their existence and properties
had been rare.
This is not surprising and is mainly due to the fact that
the vortex energy is
increasing logarithmically with the system size and it thus diverges
for systems without boundaries.

The situation has dramatically changed in the last years.
It was realized that a disc-shaped magnetic element,
with a diameter of a few hundreds of nanometers, provides
an excellent geometry for a magnetic vortex configuration.
In particular, its exchange energy is finite because the system is finite,
while its magnetostatic field vanishes everywhere except
at the vortex core,
thus making the vortex the lowest energy magnetic state.
In a few words, the interest in the vortex stems from the fact that
this is a nontrivial magnetic state which can, nevertheless, be
spontaneously created in magnetic elements (see, e.g., \cite{raabe00}).

This leads naturally to the question whether there are any further nontrivial
magnetic states which would play an important role in magnetic elements
\cite{shigeto02,castano03}.
An answer comes from a somewhat unlikely direction.
Recent experiments have shown a peculiar dynamical behaviour of vortices
and magnetic domain walls
when these are probed by external fields.
Vortices may switch their polarity under the influence
of a very weak external magnetic field of the order
of a mT \cite{waeyenberge06}.
The same switching phenomenon was observed by passing an
a.c. electical current in a magnetic disk \cite{yamada07}.
Since the polarity of the vortex contributes to its topological structure,
the switching process clearly implies a discontinuous change
of the magnetic configuration.
This is certainly a surprise since the external field
is only very weak.
The key to the phenomenon is the appearance
of vortex pairs which are spontaneously created in the vicinity
of existing vortices \cite{waeyenberge06,hertel07}.
The creation of topological excitations (vortex pairs)
by alternating external fields has been anticipated
by a collective coordinate study \cite{pokrovskii85}.

We will study vortex-antivortex pairs (VA-pairs)
and argue that these are nontrivial magnetic states
which play an important role in
dynamical phenomena in magnetic elements.
They are localized configurations - unlike a single vortex - and
they have a finite energy even when found in infinite systems.
Specifically, we will study a VA-pair where the vortex
and the antivortex have opposite polarities.
This behaves as a rotating vortex dipole.
Its topology and its dynamics are radically different than a
pair with same polarities which was shown to undergo translational
(Kelvin) motion \cite{papanicolaou98}.
Despite its nontrivial topology a rotating vortex dipole can be destroyed
by a quasi-continuous process.
In particular, no energy barrier has to be overcome in contrast to the
usual case for topological solitons.
This opens the possibility for switching mechanisms between topologically
different states in ferromagnets.
On account of a direct link between topology and dynamics
in magnetic media \cite{papanicolaou91,komineas96},
the possibility to change the topological magnetization structure leads to a
dramatic change of the magnetization dynamics
as the VA-pair is created or annihilated.
Such a pair annihilation process lies at the heart
of a counterintuitive vortex polarity switching event which was observed
in magnetic elements \cite{waeyenberge06,yamada07}.

\section{The model}
\label{sec:model}

A ferromagnet is characterized by the magnetization vector
$\bm{m} = (m_1, m_2, m_3)$ which is a function of position and time
$\bm{m} = \bm{m}(\bm{r},t)$.
It has a constant length which we choose unity for convenience:
$\bm{m}^2 = 1$.
The dynamics of the magnetization is given by the Landau-Lifshitz (LL) equation
which can be written in a rationalized form
\begin{equation}  \label{eq:lle}
\frac{\p\bm{m}}{\p t} = \bm{m} \times \bm{f}, \quad
\bm{f} \equiv \Delta\bm{m} - Q\, m_3\, \bm{\hat{e}}_3,
\end{equation}
where $\bm{\hat{e}}_3$ is the unit vector in the third magnetisation direction.
The vector $\bm{m}$ is normalized to 
the saturation magnetization $M_s$.
Distances are measured in exchange length units
$\lex \equiv \sqrt{A/(2\pi M_s^2)}$ where $A$ is the exchange constant.
The unit of time is $\tau_0 \equiv 1/(4\pi\gamma M_s)$ where $\gamma$ is the
gyromagnetic ratio.
Typical values are $\lex \sim 5{\rm nm}$
and $\tau_0 \sim  10{\rm ps}$.
We will use these values in examples given later in this paper.

In the form (\ref{eq:lle}) the LL equation has only one free parameter
which is the quality factor $Q\equiv K/(2\pi M_s^2)$, where
$K$ is an anisotropy constant.
We will typically use $Q=1$ in the following, which corresponds
to easy-plane anisotropy, unless stated otherwise.
Eq.~(\ref{eq:lle}) is associated with the energy functional
\begin{eqnarray}  \label{eq:energy}
E & = & E_e + E_a, \\
E_e & = & \half \int{(\bm{\nabla}\bm{m})^2\, d^2x},\quad
E_a = \frac{Q}{2} \int{(m_3)^2\, d^2x}, \nonumber
\end{eqnarray}
where the integration extends over an infinite plane.
We thus consider, for simplicity, a 2D infinite system.
Energies are then measured in units of $4\pi M_s^2 \lex^2$.
The anisotropy energy term in (\ref{eq:energy}) models
a magnetocrystalline anisotropy contribution which
is intrinsically present in materials, but it can also
serve as a simplified model for the magnetostatic term in thin films.
We defer discussion of a magnetostatic term 
until we derive our main results.

A magnetic vortex is an axially symmetric solution of Eq.~(\ref{eq:lle})
of the form
\begin{equation}  \label{eq:vortex0}
m_3 = \lambda\,\cos\Theta(\rho), \quad
m_1 + i\, m_2 = \sin\Theta(\rho)\, e^{i\winding (\phi - \phi_0)},
\end{equation}
where $(\rho, \phi)$ are polar coordinates, $\winding$ is an integer
called the {\it winding number}, $\lambda=\pm 1$ is the vortex {\it polarity},
and $\phi_0$ is a constant which will be called
the vortex {\it orientation}.
The magnetization angle $\Theta(\rho=0) = 0$ at the vortex center,
and $\Theta(\rho \to \infty) = \pi/2$ away from the vortex center
which corresponds to an in-plane magnetization at spatial infinity.
The winding number $\winding$ is a topological invariant.
The usual case for a vortex observed in experiments is $\winding=1$
and $psi_0=\pm \pi/2$,
while in the case $\winding=-1$ it is termed an {\it antivortex}.

A further topological invariant is defined by \cite{rajaraman}
\begin{equation}  \label{eq:topo}
\Topo \equiv \frac{1}{4\pi}\,\int{n\,d^2x},\quad
\topo \equiv \frac{1}{2}\epsilon_{\mu\nu}\,
(\p_\nu\bm{m}\times\p_\mu\bm{m})\cdot \bm{m},
\end{equation}
where $\epsilon_{\mu\nu}$ is the totally antisymmetric tensor
($\mu,\nu=1,2$), and $\topo$ is a topological density.
For a vortex of the form (\ref{eq:vortex0}) we have
$\Topo =  -1/2\,\winding\,\lambda$.
That is, for the usual vortices with $\winding=1$ we have
$\Topo = \pm 1/2$ where the sign depends on the polarity.

\begin{figure}
\epsfig{file=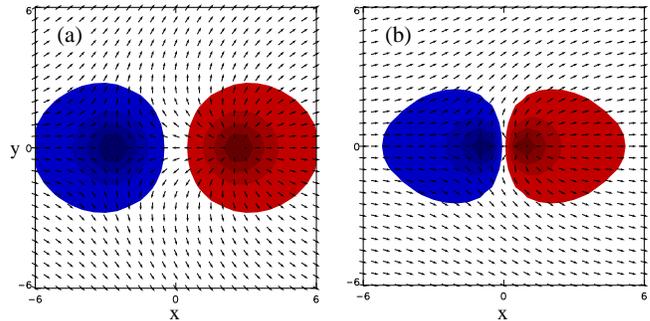,width=\linewidth}
  \caption{(colour online) VA-pairs with opposite polarities.
The vectors show the projection of $\bm{m}$ on the $(x,y)$ plane,
$m_3$ is coded in colours.
(a) The distance between vortices is relatively large ($d=2.7$)
and thus each vortex profile resembles that of an isolated one.
The pair rotates with $\omega=0.06$ ($\ell=68$).
(a) The distance between the vortices is small ($d=1.85$)
and thus the configuration is similar to (\ref{eq:twomerons}).
Rotation frequency $\omega=0.17$ ($\ell=15$).
}
\label{fig:vap}
\end{figure}

\section{A vortex-antivortex pair}
\label{sec:vap}

A convenient formulation for the description of vortices, but also
of multi-vortex states, is obtained through the stereographic variable
\begin{equation}  \label{eq:omega}
\Omega \equiv \frac{m_1 + i\, m_2}{1 + m_3}.
\end{equation}
A simple configuration for a two-vortex solution may be written as
\begin{equation}  \label{eq:twomerons}
\Omega = \frac{\bar{\zeta}+a}{\bar{\zeta}-a},\,
\qquad \zeta \equiv x+ i\, y,\;\; \bar{\zeta} \equiv x - i\,y,
\end{equation}
where $a$ is a constant which will be considered real for simplicity.
It gives the distance between vortices as well as the
size of each vortex core.
A vortex similar to (\ref{eq:vortex0}) with $\winding=1$
and $\lambda=-1$ is centered at $\zeta=a$,
an antivortex ($\winding=-1$) with opposite polarity $\lambda=1$
is centered at $\zeta=-a$.
At large distances $|\zeta| \to \infty$ we have $\Omega \to 1$.
In other words we have chosen the boundary condition
\begin{equation}  \label{eq:bc}
\bm{m} = (1,0,0) \quad {\rm as}\quad |\zeta| \to \infty,
\end{equation}
which is appropriate in the presently studied model
of an easy-plane anisotropy.
The boundary condition (\ref{eq:bc}) indicates that
- unlike in the case of a single vortex -
a VA-pair is a localised configuration, in the sense
that the magnetisation goes to the vacuum at spatial infinity.
Fig.~\ref{fig:vap} shows examples of VA-pairs, whose details
are calculated and will be discussed later in the text.

The form (\ref{eq:twomerons}) will be used as an ansatz for a
VA-pair with opposite polarities.
It belongs to a family of exact static solutions of
the Landau-Lifshitz equation
for an isotropic 2D ferromagnet ($Q=0$)
\cite{rajaraman,gross78}.
Its energy is $E = E_e = 4\pi$ for every $a$,
(due to the scale invariance of the exchange interaction \cite{rajaraman}),
and its topological charge is $\Topo=1$.
The topological density is even with respect to $x$
$\topo(x,y) = \topo(-x,y)$, and thus
precisely half of the contribution to the topological charge comes from
the vortex in the one half-plane ($x>0$) and the second half comes from the
antivortex in the second half-plane ($x < 0$).
The vortex and the antivortex are not overlapping
irrespectively of the distance between them.

In the presence of an easy-plane anisotropy term ($Q > 0$)
the profile of the vortex and the antivortex will not be similar
to that suggested by ansatz (\ref{eq:twomerons}) but it will be modified
following the scale introduced by the anisotropy.
The vortex core
size is set to $R_c \sim 1/\sqrt{Q}$, or $R_c \sim \sqrt{A/K}$
in the usual units.
Otherwise, the comments on the topological density and charge
of the VA-pair given in the previous paragraph remain valid.
Finally, the boundary condition (\ref{eq:bc}) is consistent with
easy-plane anisotropy.

\section{Vortex-antivortex dynamics}
\label{sec:dynamics}

A more detailed study of the VA-pair
requires information on its dynamics.
For this purpose we write the linear momentum associated with the
LL equation (\ref{eq:lle}) \cite{papanicolaou91}
\begin{equation}  \label{eq:linmom}
P_x = -\int{y\, \topo\, dx dy}, \quad P_y = \int{x\, \topo\, dx dy},
\end{equation}
whose components are actually the moments of the topological density $\topo$.
Since the vortex and the antivortex
have a mirror image topological density distribution
we find that $(P_x, P_y)=(0,0)$.
Formulae (\ref{eq:linmom}) give a measure of the position
of the VA-pair (which is at the origin).
The important result is that this mean position is conserved in
time since (\ref{eq:linmom}) are naturally
conserved quantities.
We thus see that the VA-pair is spontaneously pinned
at the position where it is created.
This dynamics should be contrasted to the dynamics of a VA-pair
with same polarities. The latter forms a solitary wave
which is necessarily non-static and undergoes
translational (Kelvin) motion \cite{papanicolaou98}.

Further information is obtained by considering the
angular momentum of the system \cite{papanicolaou91}
\begin{equation}  \label{eq:angmom}
\ell = \half \int{\rho^2\, \topo\, dx dy},
\end{equation}
which is also written with the aid of the topological density $\topo$,
and it is conserved within model (\ref{eq:lle}).
For a VA-pair $\ell$ is clearly nonzero
(the integrand is positive definite), and it gives
a measure of its size. The distance $d$ between
the vortex and the antivortex can be defined from
\begin{equation}  \label{eq:vapsize0}
\left(\frac{d}{2}\right)^2
   \equiv \frac{\int{\rho^2\, \topo\, dx dy}}{\int{\topo\, dx dy}}
        = \frac{\ell}{2\pi\,\Topo}\,,
\end{equation}
where we assume that the VA is centered at the origin.
Using the value $\Topo = 1$ for the vortex pair we find
\begin{equation}  \label{eq:vapsize}
   d^2 = \frac{2}{\pi}\,\ell\,.
\end{equation}
A nonzero angular momentum $\ell$ indicates that the VA
should be a rotating object.

The rotational dynamics of the VA-pair is
fully confirmed by numerical simulations of the LL Eq.~(\ref{eq:lle}).
However, the rotating pair is apparently unstable and short lived due to
radiation of energy in analogy to a rotating electric dipole.
The conservative system will be further studied here in order
to obtain some useful results.
These will then be used to understand the full dynamics of a
more realistic system including mechanisms for energy dissipation.

For a simpler numerical investigation of the main features of a rotating pair,
we consider a steady state of a rotating VA-pair
for every value of the conserved angular momentum $\ell$.
One should note that
this is a stationary point of the extended energy functional
\begin{equation}  \label{eq:eol}
F = E - \omega\, \ell,
\end{equation}
where $\omega$ is the angular frequency of rotation of the
VA-pair measured in units of $4\pi\gamma M_s$.
Using the standard scaling argument \cite{derrick64} that a stationary state
is a minimum of the functional (\ref{eq:eol}) with respect to
scaling the coordinates, one readily finds,
for a stationary state, the virial relation
\begin{equation}  \label{eq:derrick}
E_a = \omega\,\ell.
\end{equation}
The exchange energy does not appear in the above relation due to
its scale invariance in two dimensions.

A useful result is now obtained
if we assume well separated vortex and antivortex
where each of them behaves as an isolated one.
The anisotropy energy for an isolated single static vortex
in the model (\ref{eq:lle}) is $\pi/2$ \cite{komineas98}.
Substituting this number in Eq.~(\ref{eq:derrick}) we obtain
\begin{equation}  \label{eq:ol}
\omega = \frac{\pi}{\ell} = \frac{2}{d^2},
\end{equation}
where Eq.~(\ref{eq:vapsize}) was used in the second equality.
The limit of large separation between vortices was studied
in Ref.~\cite{pokrovskii85} through a collective coordinate approach
and the angular frequency (\ref{eq:ol}) was obtained.
Eq.~(\ref{eq:ol}) shows that the rotation frequency goes to zero
in the limit of large vortex separation.
As a concrete example, suppose that the vortices are a distance
$d=10\lex$ apart and $Q=1$, to obtain $\omega=0.01$, or $\omega \sim 10^9$ Hz.

We now turn to the case that the vortex and antivortex are close
and they are forming a small pair.
When the size of the VA-pair is smaller
than the length scale $1/\sqrt{Q}$ introduced by the anisotropy energy,
the latter is negligible. Then (\ref{eq:twomerons})
is exact in the area around the origin close to the vortex cores.
Away from the vortex cores, we will adopt the approximation that
the anisotropy will prevail and it will enforce the vacuum
$\bm{m} = (1,0,0)$.
In the limit $a \ll 1$ we have $E = E_e = 4\pi$ \cite{tretiakov07}
and $\Topo = 1$
due to the contribution of the origin which becomes a singular point.
The magnetic configuration becomes then fully aligned
to the $x$-axis, except for the origin as $a \to 0$.
Then, one can eliminate the
singular point and assume a fully aligned state.
In doing so, one eliminates the topological complexity of the system
and changes the topological charge $\Topo$ by unity.
Needless to say, such singular points may not exist
in magnetic materials because the lattice spacing in the solid
provides a natural length cut-off.
The finite lattice spacing is apparently responsible for the annihilation
of vanishingly small VA-pairs
and the subsequent surprising change of the topological charge
of magnetic configurations in recent experiments \cite{waeyenberge06,yamada07}.

One usually expects that a topologically nontrivial state ($\Topo \neq 0$)
cannot be deformed continuously to the ground state
of the system which has topological charge zero.
In addition, the two states are separated by an infinitely
high energy barrier.
However, in the case of the VA-pair studied
here, although the former argument is correct, no energy barrier
is encountered in the process.
This is an unusual property but it is completely explained by
the scale invariance of the exchange energy
\cite{rajaraman}.

\begin{figure}
\epsfig{file=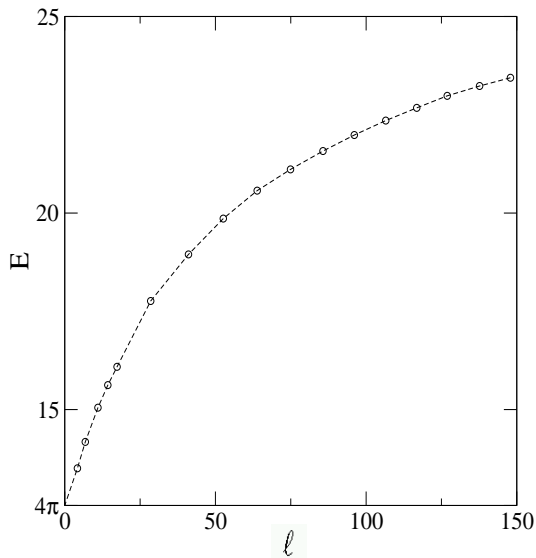,width=7cm}
  \caption{The energy $E$ of a rotating VA-pair as
a function of its angular momentum $\ell$ (for $Q=1$).
The circles indicate numerical results (the dashed line is a guide to the eye).
It can be fitted by  $E=\pi\ln(\ell/4.5)+4\pi$
for large $\ell$, while $E=0.5\ell+4\pi$ for $\ell\to 0$.
The rotation frequency is the slope of the curve.
The energy of a vanishingly small VA-pair is $4\pi$.
}
\label{fig:energy_ell}
\end{figure}

We go on to find numerically VA-pairs in steady rotation.
We set up a numerical grid which goes to spatial infinity
by using stretched coordinates $\eta \equiv a \tan x,\; \xi \equiv b \tan y$.
The choice of the constants $a,b$ depends on the size of the
studied vortex pair and they are chosen so that it is
described at sufficient resolution.
We first go to a rotating reference frame which
amounts to substituting
$\bm{f} \to \bm{f} - \omega(\delta\ell/\delta\bm{m})$
in Eq.~(\ref{eq:lle}).
We use a relaxation algorithm, i.e., we add dissipation to the LL equation
(effectively using Eq.~(\ref{eq:llge})).
We feed the algorithm with the ansatz (\ref{eq:twomerons})
as an initial condition, and this subsequently relaxes to a roughly
steady state. Our algorithm cannot give exact solutions
of the equation; a more elaborate numerical method would have to
be used for that purpose. However, using appropriate parameters for
the numerical lattice, we are able to
find approximate solutions which are almost steadily rotating states.
We thus find good approximations for the energy and the angular momentum
of the rotating state for various values of the rotation frequency $\omega$.
Fig.~\ref{fig:vap} shows rotating VA-pair solutions.
Fig.~\ref{fig:energy_ell} shows the results for the
energy as a function of the angular momentum for $Q=1$.
It is clear that the energy of a vanishingly small VA-pair (at $\ell \to 0$)
is pure exchange energy and it is equal to $4\pi$, as anticipated.
We have checked that the virial relation (\ref{eq:derrick}) is
approximately satisfied for our numerical solutions.
We note that it appeared more difficult to obtain reasonable convergence of the
algorithm for smaller values of $\ell$.

We can find the rotation frequency of a vanishingly small
VA-pair by substituting Eq.~(\ref{eq:twomerons})
in Eqs.~(\ref{eq:energy}) and (\ref{eq:angmom}) to find
\begin{equation}  \label{eq:Eaell}
E_a = \frac{Q}{2}\, \ell = Q\,
\int\limits_{S} {\frac{a^2 \rho^2}{(\rho^2 + a^2)^2}\;(2\pi\rho\, d\rho)}\,,
\end{equation}
where $S$ denotes a circular domain which includes the vortex cores.
Substituting (\ref{eq:Eaell}) in the virial relation (\ref{eq:derrick})
we find $\omega = Q/2$.
A more rigorous calculation would require an asymptotic analysis of the LL equation.
On the side of small $\ell$ in Fig.~\ref{fig:energy_ell},
we have a steep increase of $\omega$
consistent with the theoretical value $\omega(\ell\to 0) = 1/2$
at ($Q=1$).

We can give an analytic estimation of the curve in Fig.~\ref{fig:energy_ell}
for large values of the angular momentum $\ell$.
We first note that the frequency satisfies $\omega = dE/d\ell$ and it is thus given
by the slope of the curve.
For large $\ell$ we can
substitute Eq.~(\ref{eq:ol}) for $\omega(\ell)$
and obtain a differential equation which gives
\begin{equation}  \label{eq:eell}
E = \pi\, \ln(\ell/\ell_0) + 4\pi,
\end{equation}
where the constant $\ell_0$ cannot be fixed by the present calculation
(the constant $4\pi$ has been added to facilitate comparison
with Fig.~\ref{fig:energy_ell}).
The logarithmic behaviour of the energy can be obtained by considering
the energy of two widely separated vortices.
The anisotropy energy should be independent of $\ell$
while the interaction between the vortices is due to exchange.
The exchange energy
is approximately equal to $2\pi\, \ln(d/R_c)$ and this
coincides with Eq.~(\ref{eq:eell}) when we use Eq.~(\ref{eq:ol}) for large $\ell$.

We now turn to discuss more realistic systems where
dissipation is present.
The LL equation including Gilbert damping reads
\begin{equation}  \label{eq:llge}
\frac{\p\bm{m}}{\p t} = \bm{m} \times \bm{f}
 - \alpha\, \bm{m} \times (\bm{m} \times \bm{f}),
\end{equation}
where $\alpha$ is the dissipation constant.
The dissipation is necessary in order that the simulated results
are closer to the experimental, but it also helps
to rid the numerical results of any fast transients and thus
focus on the main features.
In particular, we typically consider in this paper strong enough
Gilbert damping so that it dominates the radiation effect
disussed earlier in connection to conservative dynamics.

We typically use configuration (\ref{eq:twomerons}) as an initial
state for the numerical simulation.
This is iterated in time using, typically, a dissipation constant
$\alpha\sim 0.1$ and values of $a \sim 5$.
In the initial phase
of the simulation the vortex and the antivortex cores quickly shrink 
to the size $R_c$ discussed above.
Then the vortex and the
antivortex approach each other and thus the size of the pair decreases.
This is certainly a result of dissipative dynamics (the Gilbert term
in Eq.~(\ref{eq:llge}))
as the energy (mainly the exchange part)
is obviously decreasing with the vortex pair size.
The rotational dynamics (due to the conservative term in Eq.~(\ref{eq:llge}))
is at the same time present and it becomes
prevalent for small dissipation $\alpha$.

The dynamical behaviour of a VA-pair can now be summarized as follows.
A vortex and an antivortex rotate around each other, while the
pair shrinks due to dissipation of energy.
Its energy follows approximately the curve of
Fig.~\ref{fig:energy_ell} as its size (and its angular momentum) decreases.
At vanishing size a singular point, as the one discussed earlier,
would be created. This would practically
disappear due to the discreteness of the solid.

We now turn to discuss the relevance of the above ideas
in a thin film environment.
The addition of the magnetostatic field in Eq.~(\ref{eq:lle})
or Eq.~(\ref{eq:llge})
is now necessary in order to make the model
relevant for experiments in magnets.
The qualitative picture for the dynamics of a VA-pair
will not be affected since our analysis
does not depend on the exact form
of the field $\bm{f}$ in Eq.~(\ref{eq:lle}).
The magnetostatic field will, however, modify the profile
of the pair. Unlike the exchange and anisotropy,
the magnetostatic energy is sensitive to rotations of the spin vector.
It has been pointed out \cite{pokrovskii85} that this
symmetry breaking is responsible for the resonant excitation
of vortex pair by alternating external fields.

For an ultra-thin film, we can assume that there is no variation
of $\bm{m}$ across the
film thickness, and we then expect that the magnetostatic interaction
would behave as a single-ion easy-plane anisotropy
(as for the ground state in thin films \cite{gioia97}).
This amounts to the substitution $Q \to Q+1$ in Eq.~(\ref{eq:lle}).
Our preceeding analysis is then valid
for an effective anisotropy strength $Q+1$.
In particular,
a change in the topological charge occurs through a 2D singular point.
It is interesting to note that in this particular case
no Bloch Point need be formed.
In the special case of a significant easy-axis magnetocrystalline anisotropy,
at $Q=-1$, we effectively recover the isotropic model.
The curve in Fig.~\ref{fig:energy_ell} would then be a straight line
$E(\ell) = 4\pi$.
Consequently, a VA-pair which is somehow created
in the magnet would be static. More realistically (experimentally)
one would expect that it would be slowly moving.

It would be interesting to consider the behaviour of the curve
as the thickness of the film is increased.
A full study can only be done numerically and will not be given here.
An interesting possibility is that
the long-range magnetostatic interaction
would compete with the local anisotropy interaction to
give a non-monotonic $E(\ell)$ curve, for values of $Q \approx -1$.
A minimum in the curve for small $\ell$ could then appear because
the vortex and the antivortex have
opposite polarity which may result in a low magnetostatic energy.
If such a minimum is there, we would have $\omega=0$ at that point
and thus an unusual state of a static VA-pair would be stabilized.

Numerical simulations \cite{hertel06,hertel07} show that, in a film of finite
thickness, a singular point is
created first at one of the surfaces of the film.
This should apparently be attributed to the variation of the
magnetization across the film thickness.
The VA-pair vanishes by formation and annihilation of
the singular point at successive film levels,
until the vacuum state is reached throughout the film.
A curve similar to that in Fig.~\ref{fig:energy_ell} gives now
the {\it linear} energy density across the film thickness.
The energy released when the VA-pair is annihilated is $4\pi t$ ($t$ is the
film thickness) and it is emitted
through spin waves \cite{hertel07,tretiakov07}.

At the stage when the singular point has been formed and annihilated, say,
near the top film surface, while a VA-pair still exists
in the lower film levels, we have a Bloch Point (BP)
in the film \cite{feldtkeller65}.
This lies at the interface between the region where the VA
has been annihilated and the region where this has a finite size, at the
singular point.
This is a somewhat simplified realization of the BP studied
in Ref.~\cite{doering68}.
Its energy is approximately $4\pi$ times the length of the vortex
or the antivortex line.
More important is the unusual fact that during
creation and annihilation of the BP one does not have
to overcome any energy barrier
(unlike the case discussed in Ref.~\cite{slonczewski75}), essentially
for the same reasons explained above in connection to the singular point.

\section{Conclusions}
\label{sec:conclusions}

We have studied a vortex antivortex pair where the two vortices have opposite
polarity. The pair has a nonzero topological number $\Topo = 1$,
and the vortices rotate around each other in analogy to a rotating electric dipole.
As energy is dissipated the pair shrinks and it eventually degenerates
to a singular point. Such a singular point would disappear in a solid
due to the discreteness of the lattice.
The latter process changes the topological number of the magnetic configuration.
It is surprising that no energy barrier needs to be overcome during
this quasi-continuous process.

The mechanism for changing the topological number of the magnetization
configuration makes it possible to obtain controlled
switching between topologically distinct (and therefore robust)
magnetic states.
This was achieved in the experiments of Refs.~\cite{waeyenberge06,yamada07}.
The complete mechanism, which is common in both experiments
as regards vortex dynamics,
includes a dynamical process
which involves three vortices \cite{hertel07}.
Our analysis applies only to the final part of this process,
that is, when the vortex and antivortex
with opposite polarities are relatively close together.
This is one of the important stages of the process
at which the topology of the magnetic configuration actually changes.
The presence of a third vortex does not modify our main arguments.
The full three-vortex system has
a nonvanishing topological charge and thus it is
a rotating object which is spontaneously pinned in the magnet.
A study of this configuration could be done following the
methods of this paper.

Our discussion clearly suggests a connection
between the VA-pair dynamics and the dynamics of a BP.
In particular, a rotating motion
of the magnetization appears inherent in a BP configuration.
This effect is apparently not present in the special spherically
symmetric BP solution \cite{doering68}.
It is not present either in the dissipative dynamics
of a BP discussed in the literature \cite{malozemoff}.

\vskip 5pt

\begin{acknowledgments}
I am grateful to N.R. Cooper and N. Papanicolaou for discussions
and for important comments on the manuscript.
I also gratefully acknowledge discussions with
H. Stoll, H. F\"ahnle, and O. Tchernyshyov.
\end{acknowledgments}


\end{document}